\documentclass[aps,prl,twocolumn,superscriptaddress]{revtex4}
\usepackage{graphicx}
\usepackage{color}
\usepackage{changes} 
\usepackage{xfrac} 

\newcommand{\BiTe}{Bi$_2$Te$_3$ }

\newcommand{\FeTe}{Fe$_{1.08}$Te }

\begin{document}

\title{Electronic Structure of Fe$_{1.08}$Te bulk crystals and epitaxial FeTe thin films on Bi$_2$Te$_3$}
\author{Fabian Arnold$\dag$}
\affiliation{Department of Physics and Astronomy, Interdisciplinary Nanoscience Center (iNANO), Aarhus University, 8000 Aarhus C, Denmark}
\author{Jonas Warmuth$\dag$}
\affiliation{Department of Physics, Hamburg University, Hamburg, Germany}
\author{Matteo Michiardi}
\affiliation{Department of Physics and Astronomy, Interdisciplinary Nanoscience Center (iNANO), Aarhus University, 8000 Aarhus C, Denmark}
\author{Jan Fik\'{a}\u{c}ek}
\affiliation{Institute of Physics, Academy of Sciences
	of the Czech Republic, Prague, Czech Republic}
\author{Marco Bianchi}
\affiliation{Department of Physics and Astronomy, Interdisciplinary Nanoscience Center (iNANO), Aarhus University, 8000 Aarhus C, Denmark}
\author{Jin Hu}
\author{Zhiqiang Mao }
\affiliation{Department of Physics and Engineering Physics, Tulane University, New Orleans, Los Angeles 70118, USA}
\author{Jill Miwa}
\affiliation{Department of Physics and Astronomy, Interdisciplinary Nanoscience Center (iNANO), Aarhus University, 8000 Aarhus C, Denmark}
\author{Udai Raj Singh}
\affiliation{Department of Physics, Hamburg University, Hamburg, Germany}
\author{Martin Bremholm}
\affiliation{Department of Chemistry, University of Aarhus, Aarhus, Denmark}
\author{Roland Wiesendanger}
\affiliation{Department of Physics, Hamburg University, Hamburg, Germany}
\author{Jan Honolka}
\affiliation{Institute of Physics, Academy of Sciences of the Czech Republic, Prague, Czech Republic} 
\author{Tim Wehling}
 \affiliation{Institut f\"{u}r
 	Theoretische Physik, Universit\"{a}t Bremen, Bremen, Germany} 
 \author{Jens Wiebe}
 \affiliation{Department of Physics, Hamburg University, Hamburg, Germany}
\author{Philip Hofmann}

\email{philip@phys.au.dk}
\affiliation{Department of Physics and Astronomy, Interdisciplinary Nanoscience Center (iNANO), Aarhus University, 8000 Aarhus C, Denmark}
\date{\today}
$\dag$: Both authors contributed equally to this work
\begin{abstract}
The electronic structure of thin films of FeTe grown on Bi$_2$Te$_3$ is investigated using angle-resolved photoemission spectroscopy, scanning tunneling microscopy and first principles calculations. As a comparison, data from cleaved bulk \FeTe taken under the same experimental conditions is also presented. Due to the substrate  and thin film symmetry, FeTe thin films grow on Bi$_2$Te$_3$ in  three  domains, rotated by 0$^{\circ}$, 120$^{\circ}$, and 240$^{\circ}$. This results in a superposition of photoemission intensity from the domains, complicating the analysis. However, by combining bulk and thin film data, it is possible to partly disentangle the contributions from three domains. We find a close similarity between thin film and bulk electronic structure and an overall good agreement with first principles calculations, assuming a p-doping shift of 65~meV for the bulk and a renormalization factor of around 2.  By tracking the change of substrate electronic structure upon film growth, we find indications of an electron transfer from the FeTe film to the substrate. No significant change of the film's electronic structure or doping is observed when alkali atoms are dosed onto the surface. This is ascribed to the film's high density of states at the Fermi energy. This behavior is also supported by the ab-initio calculations.
\end{abstract}
\maketitle

\section{Introduction}


The discovery of unconventional superconductivity in iron-based superconductors (SCs)  has attracted considerable attention in the physics community in  recent years, especially the notion of high temperature superconductivity in these systems \cite{Takahashi:2008aa,Paglione:2010aa}.  Iron chalcogenides are particularly interesting as a model system for iron-based SCs because of their simple crystal structure and because of the recent discovery of high-temperature interfacial superconductivity in FeSe single layers on SrTiO$_3$ \cite{Ge:2015,Liu:2015ae}. The key ingredients for  high temperature superconductivity in  iron chalcogenides are so far unknown and it is  discussed if  phonons \cite{Lee:2014}, magnetic fluctuations \cite{Fang:2008},  two-dimensional effects \cite{Seo:2016,Song:2011} or charge doping \cite{Song:2016,Zhang:2014} are responsible.  

Thin layers of iron chalcogenides are particularly interesting, not only because of the possibility of high-temperature interfacial superconductivity  \cite{Ge:2015} but also because of the ease of doping, either by the interface or by chemical doping \cite{Miyata:2015}. So far, the focus has been on thin films of FeSe, a system which is also superconducting in the bulk \cite{Hsu:2008aa,Eich:2016,Singh2016}, while little work has been performed on FeTe layers, a compound which is not a bulk SC. On the other hand,  Fermi surface topology suggests that, in the regime of chemically doping to the optimum carrier density,  FeTe has stronger superconducting  pairing than FeSe and, if the same mechanisms apply in both materials, it is predicted that  FeTe  has a higher T$_C$ than FeSe \cite{Subedi:2008}. Moreover, thin film FeTe grown on the topological insulator Bi$_2$Te$_3$ \cite{He:2014ab,HE201435} shows an energy gap indicative of superconductivity below 6~K \cite{Manna:2017}, in contrast to the non-superconducting bulk phase. This is particularly interesting, since interfaces of s-wave superconductors with topological insulators are predicted to resemble spin-less p$_x$-p$_y$ superconductors that may support Majorana bound states in vortices \cite{Liang2007}.
 
Here we present a study of thin FeTe layers grown on  Bi$_2$Te$_3$ \cite{Manna:2017,Hanke:2017} by a combination of scanning tunneling microscopy (STM), angle-resolved photoemission spectroscopy (ARPES) and first principle calculations. Since thin film FeTe grows in three domains on the hexagonal substrate, the domain-averaged electronic structure measured by ARPES is difficult to disentangle. However, the results can be understood using ARPES data from high-quality bulk Fe$_{1.08}$Te single crystals taken under the same experimental conditions, which mostly confirm previous ARPES measurements of bulk single crystals \cite{Xia:2009, Zhang:2010aa,Liu:2013}. Finally, we investigate the possibility of doping the thin film FeTe by alkali adsorption.

\section{Experimental and Theoretical Methods}

\subsection{Thin film preparation}
Bi$_2$Te$_3$ crystals synthesized using a Stockbarger method as described in Ref. \cite{Manna:2017}, were cleaved in ultra-high vacuum (UHV) to obtain a clean surface. Afterwards, Fe was deposited from an electron beam evaporator onto the Bi$_2$Te$_3$ surface at room temperature and subsequently annealed for 30~min at 588~K. In this growth mode, part of the Te of the topmost quintuple layer (QL) consisting of the five atomic sublayers (Te-Bi-Te-Bi-Te) reacts with the deposited Fe resulting in the formation of FeTe layers of different thickness at the surface \cite{Manna:2017,Hanke:2017}. The topmost \BiTe QLs are thereby partly fractured, as shown for annealed \BiTe in the supplementary material of \cite{Schouteden2016}. This results in FeTe layers embedded into the topmost \BiTe QLs. The same growth behavior is observed for the growth of FeSe on Bi$_2$Se$_3$ \cite{Singh2016, Eich:2016}.

Rb atom deposition on the thin films has been carried out by resistance heating of a (SAES) alkali metal dispenser facing the sample surface at a sample temperature of 90~K. The coverage was calibrated by the ratio of the Rb/Te core level photoemission intensities.

\subsection{Bulk crystal growth and surface preparation}
High-quality Fe$_{1+y}$Te single crystals were synthesized using the flux method \cite{Liu:2009aa} where the excess Fe ratio $y$ was kept as low as possible. The measured composition of the crystals using single crystal X-ray diffraction resulted in $y\approx8$\%.
Fe$_{1.08}$Te crystals were cleaved in UHV. Note that Fe$_{1+x}$Te exhibits a magneto-structural phase transition at a N\'{e}el  temperature of T$_N=$60-70~K \cite{Chen:2009aa,Bao:2009ab}, depending on the amount of excess Fe, separating the paramagnetic tetragonal ($>$T$_N$) and the antiferromagnetic monoclinic ($<$T$_N$) phase \cite{Liu2015}. Angle-resolved photoemission spectroscopy (ARPES) and scanning tunneling microscopy (STM)  data  were acquired at   temperatures of 90~K and  32~K, respectively, i.e. above and below $T_N$. However,  in terms of  atomic resolution STM data, which are acquired with a non-magnetic tip,  no difference is observable between the paramagnetic tetragonal and antiferromagnetic-monoclinic phase, due to the very similar lattice constants of the two  phases.

\subsection{STM}
The STM measurements were performed with a homebuilt variable temperature STM \cite{Eich2014, Warmuth2016} located in the UHV system used for the FeTe thin film preparation. The tip and sample were cooled down to 32 K. The presented STM topographies were recorded in constant-current mode, using a tunneling current ($I_t$) and sample bias ($V_s$).

\subsection{Density Function Theory}
Density function theory (DFT) calculations were performed within the generalized gradient approximation (GGA) \cite{PBE1996} using the Vienna Ab Initio Simulation Package (VASP) \cite{Kresse:PP_VASP} and the projector augmented wave (PAW) \cite{Bloechl:PAW1994, Kresse:PAW_VASP} basis sets. We considered the tetragonal phase of stoichiometric FeTe with the lattice parameters from Ref. \onlinecite{Ciechan_JPCM2014}, i.e. $a=3.8$\AA\,, $c=6.5$\AA\, and the height of the Te atoms below/above the Fe planes being $h_{\rm Te}=1.75$\AA . The Brillouin zone integrations were performed with the tetrahedron method on $k$-meshes of size $17\times 17\times 17$.

Clearly, electronic correlation effects affect electronic spectral functions in FeTe as can be seen e.g. from dynamical mean field theory simulations of this material \cite{Yin:2011,Yin:2012}. Furthermore, the shape of the Fermi surface and the DFT band structures are quite strongly lattice geometry dependent in FeTe \cite{Subedi:2008,Ciechan_JPCM2014,Savrasov2009}, which is similar to the case of FeSe \cite{Watson:2017}. Here, the DFT calculations serve as qualitative guide to the ARPES spectra, and we ``fit'' them to the bulk ARPES data by choosing the above mentioned geometry, by applying a renormalization factor of $\approx2$ to the bandwidth and by shifting the Fermi level.

\subsection{ARPES}
ARPES data and low-energy electron diffraction (LEED) images were acquired at the SGM3 beamline of the ASTRID2 synchrotron light source \cite{Hoffmann:2004} at a temperature of 90 K.  The energy and angular resolution were better than 25~meV and 0.2$^{\circ}$, respectively. The sample orientation was determined by LEED.

\section{Results}
\subsection{Electronic structure of bulk crystals}

\begin{figure}[b!]
	\begin{center}
		\includegraphics[width=1.0\linewidth]{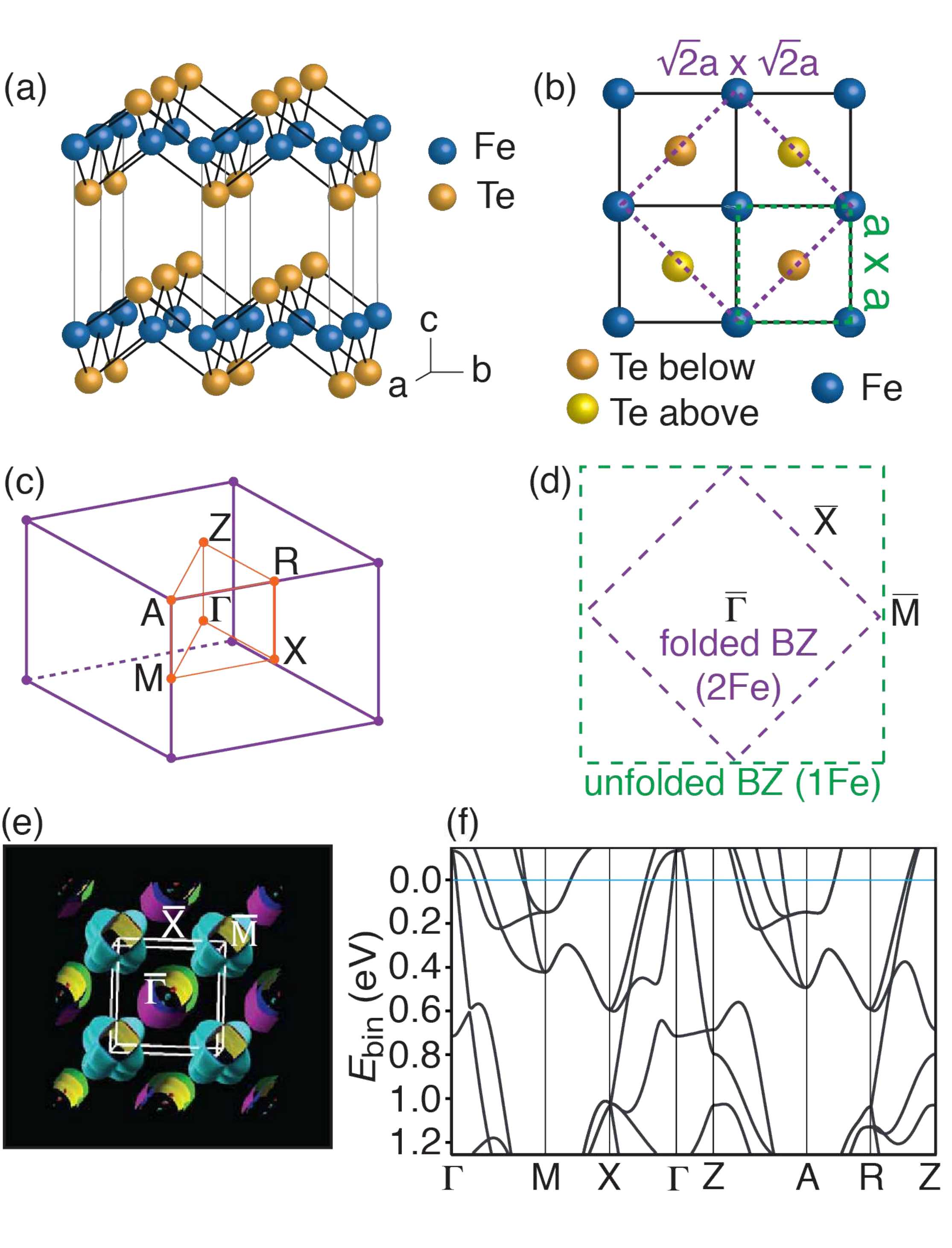}
		\caption{(Color online) (a) Side view of the PbO-crystal structure of FeTe.  (b) Top view of a single layer of  FeTe, indicating Te below and above the Fe plane.  Dashed green and dashed purple squares represent the unfolded (1-Fe) and folded (2-Fe) unit cells, respectively. (c) Bulk BZ. (d)  Unfolded (dashed green) and folded (dashed purple) BZ for a single layer of FeTe. (e) Calculated bulk Fermi surface. (f) Calculated bulk band structure along high symmetry directions. The light blue line indicates the Fermi level of the  calculation.  }
			\label{theoryFeTe}
	\end{center}
\end{figure}

We start with a presentation of the experimental and theoretical results for bulk Fe$_{1.08}$Te, as this will serve as an introduction into the properties of the materials and is necessary to understand the properties of the thin films. 
The FeTe structural polymorph of interest here has the layered PbO-crystal structure (space group P4/nmm), two layers of which are shown in Fig.  \ref{theoryFeTe}(a). The bulk Brillouin zone (BZ) is shown in Fig. \ref{theoryFeTe}(c). The three-dimensional character of the crystal structure is frequently ignored and the system is treated as essentially two-dimensional.  For the two-dimensional lattice, two different choices of the unit cell are common (Fig.  \ref{theoryFeTe}(b)): The crystallographically correct one contains two Fe and two Te atoms, one above and one below the plane of Fe atoms. The corresponding BZ is often referred to as the ``folded'' or ``2-Fe'' BZ . A smaller choice of unit cell contains only one Fe atom; it is the same unit cell that would be obtained ignoring the Te atoms altogether. The corresponding BZ is called the ``unfolded'' or ``1-Fe'' BZ (Fig.  \ref{theoryFeTe}(d)). In the remainder of the paper, we only use the larger unit cell and the folded BZ. Moreover,  it turns out that the view of FeTe as nearly two-dimensional is inadequate for the description of the electronic structure and we will take the fully three-dimensional band structure into account. Thus, instead of denoting the centre, corner and edge high symmetry points as $\Gamma$, $M$ and $X$, we label these as points in the (projected) \emph{surface} BZ and denote them by  $\bar{\Gamma}$, $\bar{M}$ and $\bar{X}$ corresponding to the bulk $\Gamma-Z$, $M-A$ and $X-R$ lines, respectively.
 
Fig. \ref{theoryFeTe}(e) and (f) show the calculated bulk Fermi surface and band structure, respectively. By comparison to the band structure calculation, the Fermi surface elements can be identified as hole pockets around the zone centre and electron pockets around the zone corner. It is clearly seen that some Fermi surface elements are ``warped'' cylinders, suggesting a non-negligible dispersion of the band structure along $k_z$. The most extreme case of this is the innermost hole pocket around $\Gamma$ which forms a closed feature in the $k_z$ direction. Other elements are merely cylindrical, consistent with a quasi two-dimensional character of the material.  The results agree well with earlier band structure calculations for this system employing the same lattice parameters \cite{Ciechan_JPCM2014}. Note that the Fermi level in the calculation, indicated by the light blue line in Fig. \ref{theoryFeTe}(f), does not necessarily have to coincide with the Fermi level in the experiment. 

\begin{figure}[b!]
	\begin{center}
		\includegraphics[width=1.0\linewidth]{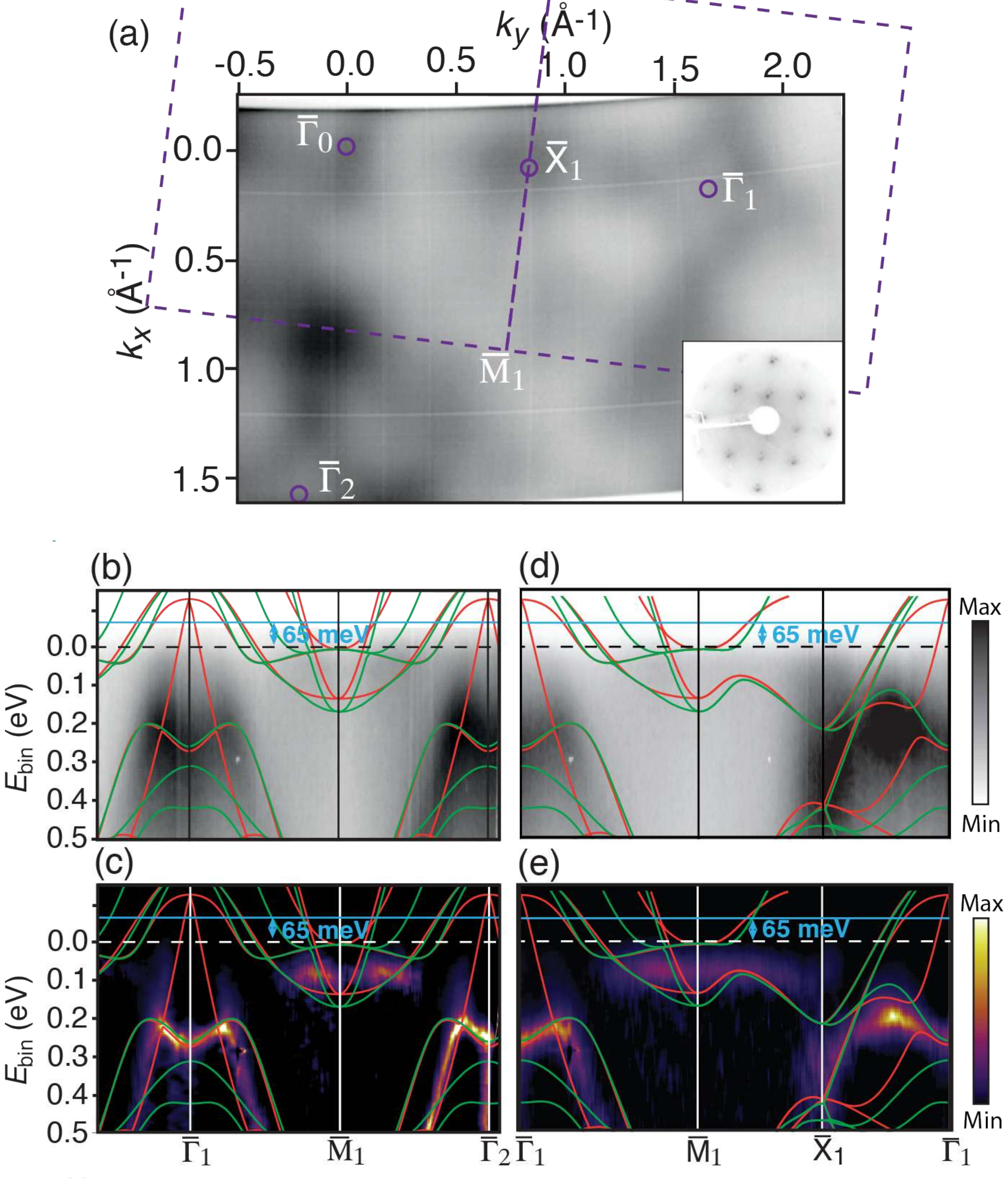}
		\caption{(Color online) ARPES results from bulk Fe$_{1.08}$Te. (a) Constant energy contour acquired at $h{\nu} = 65$~eV by integrating the spectral weight in a 120~meV window around the Fermi energy.  The purple dashed squares  visualize the projected Fe$_{1.08}$Te first and second bulk BZs with their high symmetry points (marked by small purple circles).  The inset shows  the LEED pattern used to determine the sample orientation (E$_{kin}$=128~eV).  (b) Photoemission intensity  along the 	$\bar{\Gamma}_1$-$\bar{M}_1$-$\bar{\Gamma}_2$
			directions and  (c) the corresponding curvature  plot. The solid red and  green lines are the first principles calculations for the dispersion along $\Gamma$-$M$-$X$ and $Z$-$A$-$R$, respectively. The calculations have been renormalized by a factor of $\approx 2$ and shifted to align best with the experimental data. The light  blue line indicates the Fermi level of the first principle  calculations and the dashed black and white lines the Fermi level of the photoemission and curvature data, respectively. (d), (e) Photoemission intensity and curvature  along the 	$\bar{\Gamma}_1$-$\bar{M}_1$-$\bar{X}_1$-$\bar{\Gamma}_1$	directions. }. 
		
		\label{bulkExp}
	\end{center}
	\end{figure}

Prior to collecting data with ARPES, the sample was aligned using LEED. The inset of Fig. \ref{bulkExp}(a) is a typical diffraction pattern showing the square of the surface reciprocal lattice. Note that the most intense spots are accompanied by weaker satellites that are ascribed to the existence of several slightly differently oriented crystallites within the probing area of the electron beam (diameter of $\approx$1~mm). When ARPES data are collected, care was taken to centre the light spot (horizontal size of $\approx$150~$\mu$m) on a single of these crystallites and to remain at this position during the angle scan.

The ARPES photoemission intensity  integrated in a 120~meV window around the Fermi energy and along high symmetry directions in the surface BZ is shown in Fig. \ref{bulkExp}. We choose to number equivalent high-symmetry points in reciprocal space. $\bar{\Gamma}_0$ stands for normal emission whereas the other $\bar{\Gamma}_n$ points lie in neighboring surface BZs. The ARPES features of (almost) stoichiometric FeTe are very broad at this temperature with no sharp quasiparticle peaks near the Fermi level, consistent with earlier findings from this system \cite{Xia:2009, Zhang:2010aa,Liu:2013}. Note that this situation is drastically different for many other iron-based superconductors for which sharp quasiparticle peaks can be observed \cite{Ye:2014aa}, even for FeTe when 10\% of the Te atoms are substituted by Se. A possible origin for this markedly strong difference between FeTe and other iron based superconductors are strong correlations and proximity of FeTe to an orbitally selective Mott transition \cite{Yin:2011}. Furthermore, strong matrix element effects permit the observation of certain bands only for specific polarization conditions \cite{Xia:2009,Starowicz:2013aa}. Hence not all DFT-predicted bands can be expected to be observed for one fixed light polarization here. An interesting aspect is that the observable features also strongly dependent on the BZ probed, as seen especially well in the Fermi contour of Fig. \ref{bulkExp}(a).  In the first BZ, one observes broad features around the $\bar{\Gamma}_0$ and $\bar{X}$ points, whereas more details are visible in the neighboring zones, especially around $\bar{\Gamma}_1$, $\bar{\Gamma}_2$ and $\bar{M}_1$.  As the band topography is clearer in the second BZ we perform the following analysis there. 

Detailed plots of the photoemission intensity along high-symmetry directions are shown in Fig. \ref{bulkExp}(b) and (d) with the corresponding curvature plots \cite{Zhang:2011aa} in  Fig. \ref{bulkExp}(c) and (e). 
The calculations of Fig. \ref{theoryFeTe}(f) have been superimposed on the data. When taking the three-dimensional nature of the electronic structure into account, it has to be emphasized that the component of the crystal momentum perpendicular to the surface, $k_z$, is not conserved in the photoemission process and it is therefore not known, to which value of $k_z$ the observed dispersion corresponds. Moreover, $k_z$ of the initial state can not be expected to be constant in a scan covering several surface BZs at a constant photon energy. This is not a severe limitation for many of the electronic bands that are nearly two-dimensional. However,  in order to take this possibility into account, we show a  plot of the calculated  band structure from both the middle plane of the bulk BZ (along $\Gamma$-M-X, red solid line) and along the BZ boundary (along Z-A-R, green solid line). By comparing these dispersions, it becomes immediately clear which bands show a pronounced three-dimensional character with a strong dispersion perpendicular to the FeTe planes: Such three-dimensional character is reflected in significant differences between the red and the green bands. A clearly three-dimensional feature is the innermost hole pocket around the $\bar{\Gamma}$ point that is only visible as a red band. This behavior is consistent with the $d_{xz}$/$d_{yz}$ orbital character of this band \cite{Starowicz:2013aa}. Other features, on the other hand, are nearly two-dimensional, as seen by the very similar dispersion in the red and green bands,  for example those forming the electron pockets around  $\bar{M}$  and also the camel-back shaped band near $\bar{\Gamma}$ with a binding energy of $\approx$250~meV. These two-dimensional bands are expected to appear narrower in the ARPES data because the uncertainty in $k_z$, and the accompanying broadening, does not affect them. They are thus particularly useful for aligning the calculated and measured band structures. Here, the best visual agreement is obtained when the calculations are renormalized by a factor of $\approx$2 and shifted by 65~meV. The need of a similar renormalization of  DFT-calculated bands and ARPES results has been observed before and was assigned to correlation effects \cite{Xia:2009}. The required shift is in line with a net hole doping of our \FeTe crystals, while, however, many-body renormalizations are expected to be orbitally dependent \cite{Yin:2011} and contributions from such correlation effects could also result in an effective shift of certain spectral features.

Some features predicted by the calculations are directly seen in the raw ARPES data, for example the band forming a hole pocket state around $\bar{\Gamma}_1$. Indeed, this hole pocket is also  seen in photoemission intensity at the Fermi level in Fig. \ref{bulkExp}(a), whereas it cannot be observed very clearly around normal emission at  $\bar{\Gamma}_0$. This does not only illustrate the role of matrix elements in this compound but also the importance of using the crystallographically correct ``folded'' BZ, as a unit cell based on merely the iron atoms could not necessarily be expected to show a hole pocket at this position. Other structures in the data can only be identified in the curvature plots, e.g. the bottom of the bands forming the electron pockets around $\bar{M}$. Closer to the Fermi energy, however, the bands forming these electron pockets are too weak to be observed. The strong photoemission intensity along the $\bar{X}_1-\bar{\Gamma}_1$ direction is probably due to multiple bands.
 
The photoemission intensity at the Fermi level shows a pronounced  feature near the $\bar{X}$ points, which does not correspond to any band of the calculated Fermi surface. This has also been observed in other ARPES measurements of FeTe \cite{Xia:2009,Liu:2013}, and also in Ref. \cite{Zhang:2010aa} if the surface BZ in that work is rotated by $45^{\circ}$ (as already pointed out in Ref. \cite{Liu:2013}).  The dispersion in Fig. \ref{bulkExp}(d) suggests that this feature could stem from the very intense and broad band at higher binding energy.

\subsection{Atomic and electronic structure of thin films}

\begin{figure}[b!]
 	\begin{center}
 		\includegraphics[width=1.0\linewidth]{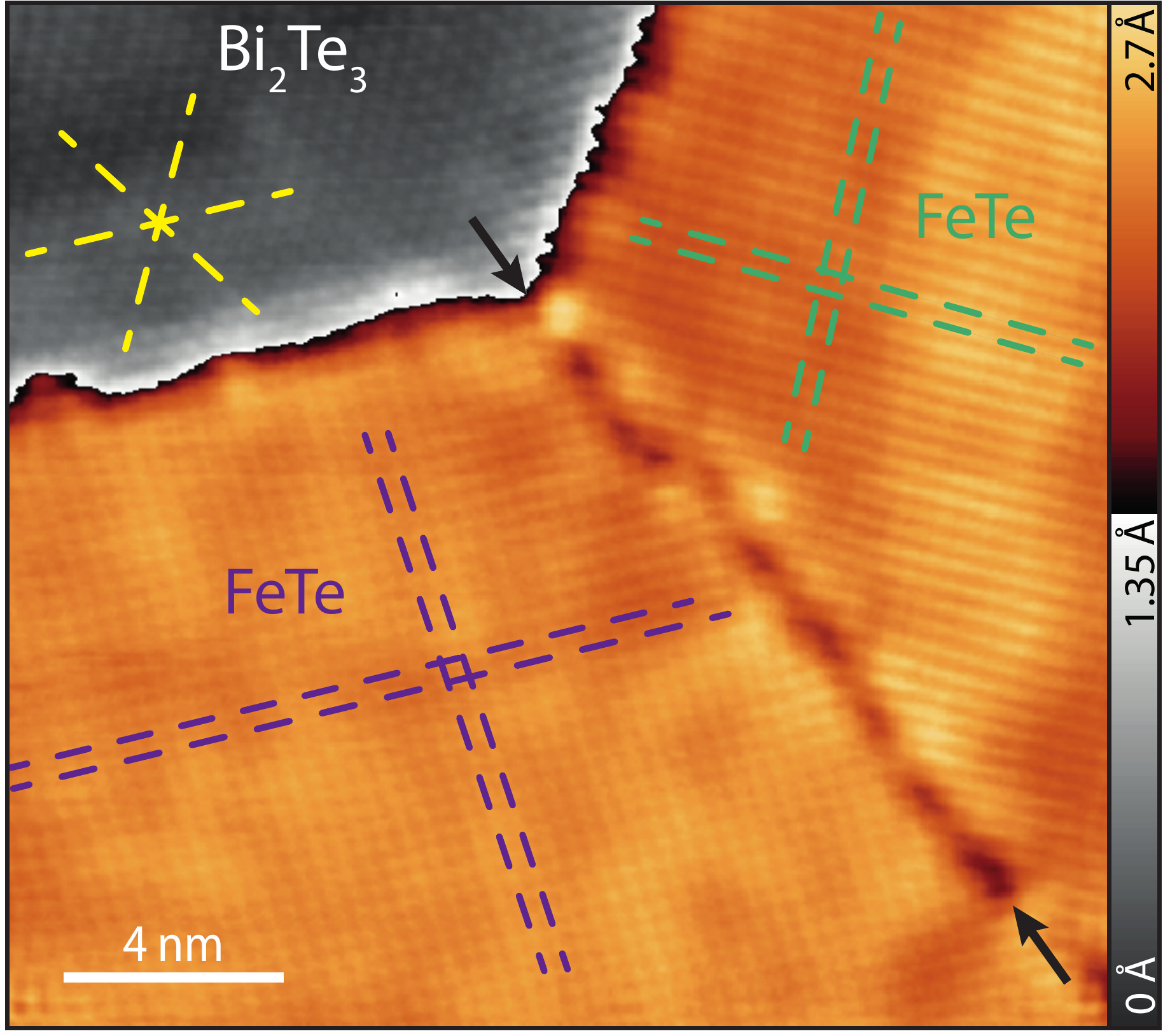}
 		\caption{(Color online)  Constant current STM image displaying bare \BiTe (hexagonal lattice) and two different rotational FeTe domains (square lattice) with atomic resolution. The color scale is selected to separate the surface area of \BiTe (gray scale) from that of FeTe (colored). Black arrows point to the grain boundary between the two rotational domains. The orientation of the atomic lattices in each surface area is marked by dashed lines (yellow for Bi$_2$Te$_3$, purple and blue for FeTe). Tunneling parameters: V$_s=$-200~mV, I$_t=$300~pA.}
		 
 		\label{thinfilmSTMb}
 	\end{center}
 	
 \end{figure}

Using Bi$_2$Te$_3$ and Bi$_2$Se$_3$ as substrates for the growth of high-quality FeTe and FeSe films, respectively, is a well-established approach  \cite{Singh2016,Eich:2016,Hanke:2017,Kamlapure:2017aa,Manna:2017} but the combination of the rectangular FeTe and the hexagonal Bi$_2$Te$_3$ surface lattice gives rise to three different rotational domains of the thin film with respect to the substrate. These different rotational domains can be studied individually by STM, as seen in Fig. \ref{thinfilmSTMb}. A surface averaging technique such as ARPES probes all rotational domains within the beam spot  simultaneously,  rendering the interpretation of the data more difficult. This is a known effect for similar systems, such as FeSe layers that are grown on graphene on SiC \cite{Song:2011}. In the following, we show how to partly disentangle the photoemission data from different domains, guided by information from bulk single crystals obtained above, such that a comparison to calculations becomes feasible. 

 \begin{figure}[b!]
 	\begin{center}
 		\includegraphics[width=1.0\linewidth]{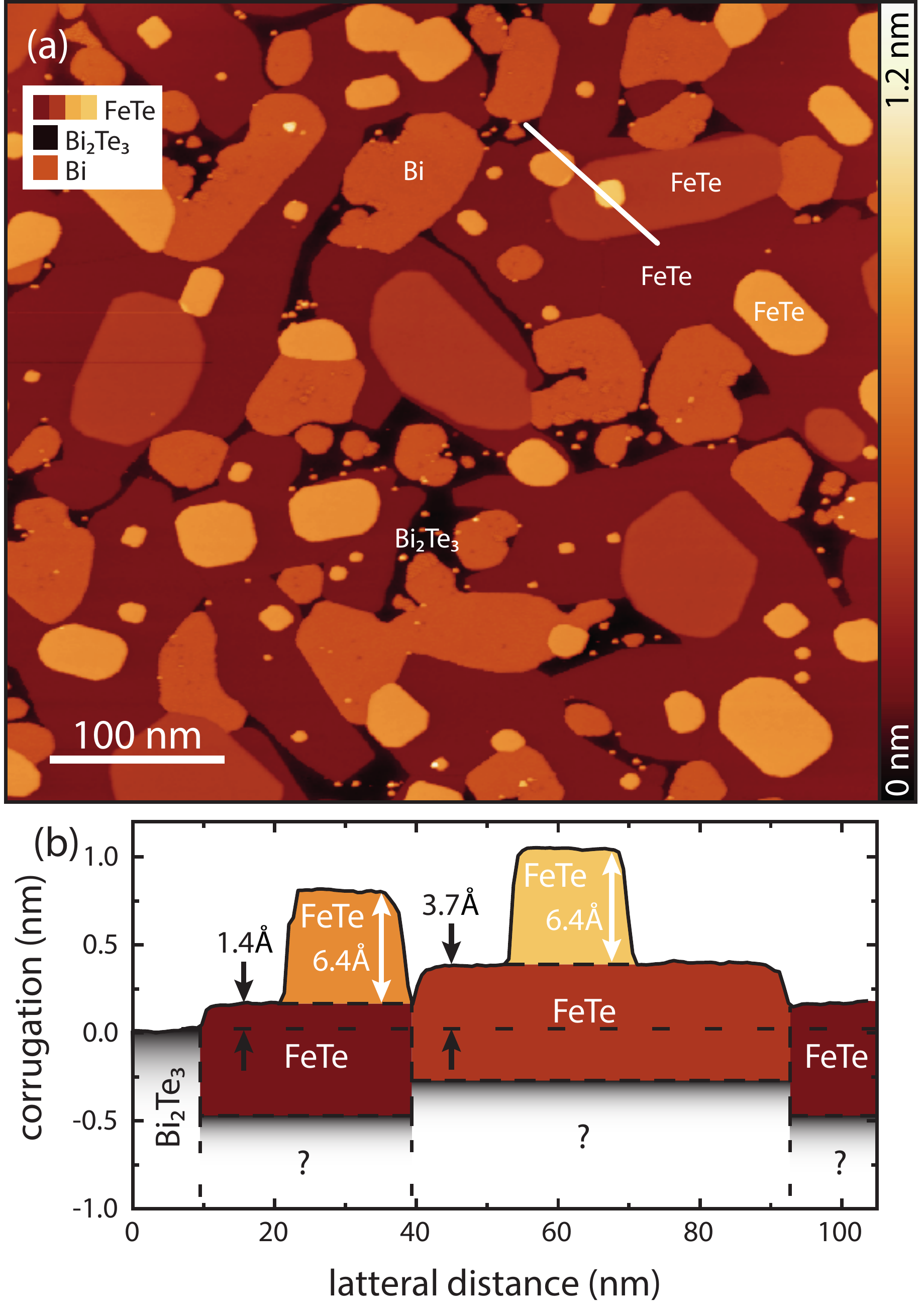}
 		\caption{(Color online) (a) Constant current STM image of the FeTe film on Bi$_2$Te$_3$ sample investigated by ARPES. The line profile plotted in (b) indicates the height information along the solid line in (a). The FeTe layers are illustrated by a color corresponding to the color of the respective surface area in (a). Tunneling parameters: (a) V$_s=$-400~mV, I$_t=$100~pA.}
		 
 		\label{thinfilmSTMa}
 	\end{center}
 	
 \end{figure}

Fig. \ref{thinfilmSTMa} shows the surface structure of the FeTe/\BiTe system probed by ARPES as seen by STM. Fig. \ref{thinfilmSTMa}(a) is an overview scan in which FeTe islands of varying apparent height are surrounded by the \BiTe substrate. The height profile along the solid line in (a) is shown in Fig. \ref{thinfilmSTMa}(b) and indicates the different apparent heights of the FeTe layers. The FeTe layers having apparent heights of 1.4$\pm$0.1~\r{A} and 3.7$\pm$0.1~\r{A} above the \BiTe surface, which is much shallower than the thickness of a single UC of bulk FeTe of 6.25 \r{A} \cite{Li2009}, are most probably embedded into the topmost \BiTe QL, as illustrated by the sketch of the layers in Fig. \ref{thinfilmSTMa}(b). This strongly indicates that sublayers of the topmost \BiTe QL are incorporated into the FeTe layer during the growth process. The irregularly shaped rough islands indicated by Bi in Fig. \ref{thinfilmSTMa}(a) exhibit an apparent height of $4.8\pm 0.1$~\r{A}. Considering the matching apparent height $4.7$~\r{A} of a single Bi bilayer on \BiTe \cite{Yang:2012aa}, these islands are presumably composed of remaining Bi from the growth reaction. The FeTe layers grown ontop of the embedded FeTe layers have an apparent height of $6.4\pm 0.1$~\r{A} in close agreement to the thickness of a single UC of bulk FeTe. While the overall FeTe thin film covers $\sim70\%$, these ontop FeTe layers cover $\sim10\%$ of the total surface area. However, we cannot exclude that there are additional layers of FeTe hidden below the embedded FeTe layers.

Fig. \ref{thinfilmSTMb} shows an atomically resolved image of the bare \BiTe and two neighboring embedded FeTe islands, revealing their local atomic structure. The orientation of the FeTe islands can be readily identified by the atomic rows of top-layer atoms. The direction of the rows is indicated by the dashed lines for the two domains in Fig. \ref{thinfilmSTMb}. Due to the simultaneously imaged atomic structure of the surrounding \BiTe (yellow dashed lines), the orientation of the domains with respect to the substrate is also evident. We can therefore conclude, that there are three rotational domains of the FeTe layers grown on the \BiTe whose lattices are rotated by 120$^{\circ}$ with respect to each other. 

\begin{figure}[b!]
	\begin{center}
		\includegraphics[width=0.8\linewidth]{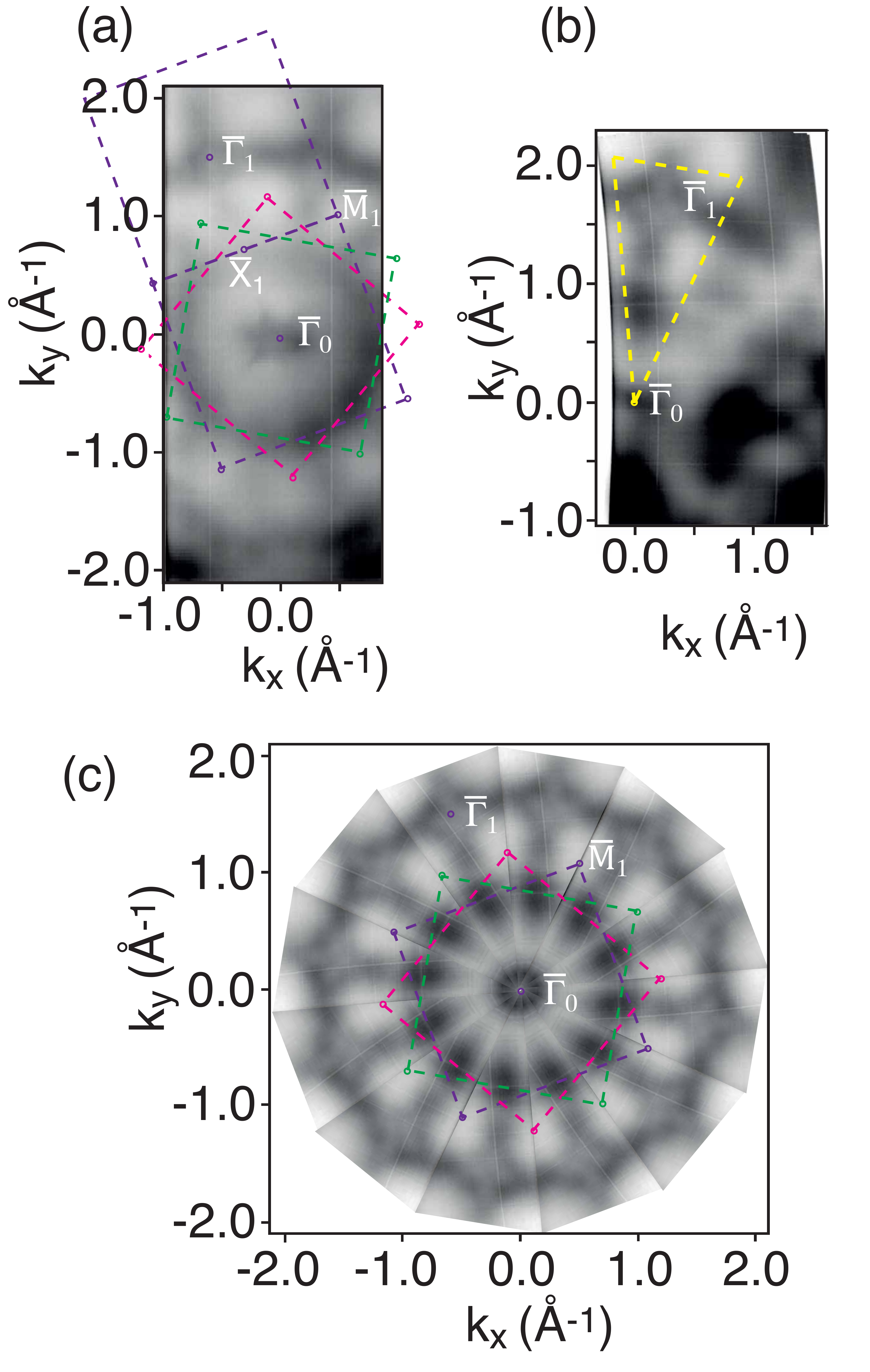}
		\caption{(Color online) (a)  Constant energy contour of thin film FeTe acquired at $h{\nu} = 65$~eV by integrating the spectral weight in a 120~meV window around the Fermi energy, revealing  the orientations of thin film FeTe in three domains (120$^\circ$ to each other). The BZ of each orientation  is represented by a pink,  purple  or  green dashed  square, respectively, and with $\bar{\Gamma}_0$, $\bar{M}_1$ and $\bar{\Gamma}_1$ indicated for the purple BZ. The high symmetry points of the purple BZ are marked by small purple circles. (b),(c) Construction of the thin film FS: (b)  Constant energy contour of bulk \FeTe from Fig. \ref{bulkExp}(a). The yellow dashed 30$^\circ$ wedge indicates the area of the bulk FS used for constructing the thin film FS. (c) With the marked area from (b) the thin film FS is constructed by rotating this wedge in steps of 30$^\circ$.}
		
		\label{FeTe_TF_bulk_FS_comp}
	\end{center}
	
\end{figure}

Fig. \ref{FeTe_TF_bulk_FS_comp}(a) shows the photoemission intensity at the Fermi energy of thin film FeTe on Bi$_2$Te$_3$, obtained with a photon energy of $h\nu=  65$~eV and integrated in a 120~meV window around the Fermi level. The main features observed are an enhanced intensity at normal emission ($\bar{\Gamma}_0$), a ring with a radius of $\approx$0.8~\AA$^{-1}$ around the origin and a second ring at $\approx$1.7~\AA$^{-1}$ containing cross-like features strongly reminiscent of the photoemission intensity around $\bar{\Gamma}_1$ in Fig. \ref{bulkExp}(a). 
As we shall see below, the observed structures can all be understood in terms of the electronic structure of bulk Fe$_{1.08}$Te.

The presence of rotational domains  is expected to affect the observed photoemission intensity away from the $\bar{\Gamma}_0$ point because of the incoherent superposition of photoemission intensity from these domains. The FeTe BZs corresponding to the domain orientations are indicated by the dashed lines in Fig.  \ref{FeTe_TF_bulk_FS_comp}(a). The observed Fermi level intensity map from the thin film can be constructed using data from the bulk sample: Fig.\ref{FeTe_TF_bulk_FS_comp}(b) shows the corresponding photoemission intensity from bulk Fe$_{1.08}$Te, acquired under the same conditions (same data set as in Fig.  \ref{bulkExp}(a) with a partly saturated greyscale). The main features of the bulk Fermi contour are all included into the yellow 30$^\circ$ wedge. Rotating this wedge in steps of 30$^\circ$, corresponding to the symmetry-equivalent directions of the differently colored surface BZs in Fig. \ref{FeTe_TF_bulk_FS_comp}(a), yields the Fermi contour in Fig. \ref{FeTe_TF_bulk_FS_comp}(c) which is very similar to the actual thin film result in Fig. \ref{FeTe_TF_bulk_FS_comp}(a). 

Upon closer inspection, however, some differences between the bulk and thin film Fermi contour can be noticed. Especially clear are changes around the $\bar{\Gamma}_1$ point, where the cross-like feature shows a clear ``hole'' in the middle for the bulk crystal, consistent with the elliptical Fermi contour expected from the presence of the predicted hole pocket. For the thin film Fermi contour this ``hole'' is still observable but it is noticeably smaller. The dispersion of the hole pockets in Fig. \ref{theoryFeTe} indicates that a smaller ``hole'' would correspond to a smaller amount of $p$-doping in the film. However, in view of the material's three-dimensional character, the fact that only some of the Fermi surface features can be observed and the intrinsically broad bands, a quantitative determination of the doping is not possible. Finally, strain in the film could also play a role \cite{Wu:2016aa}. 

\begin{figure}[b!]
	\begin{center}
		\includegraphics[width=1.0\linewidth]{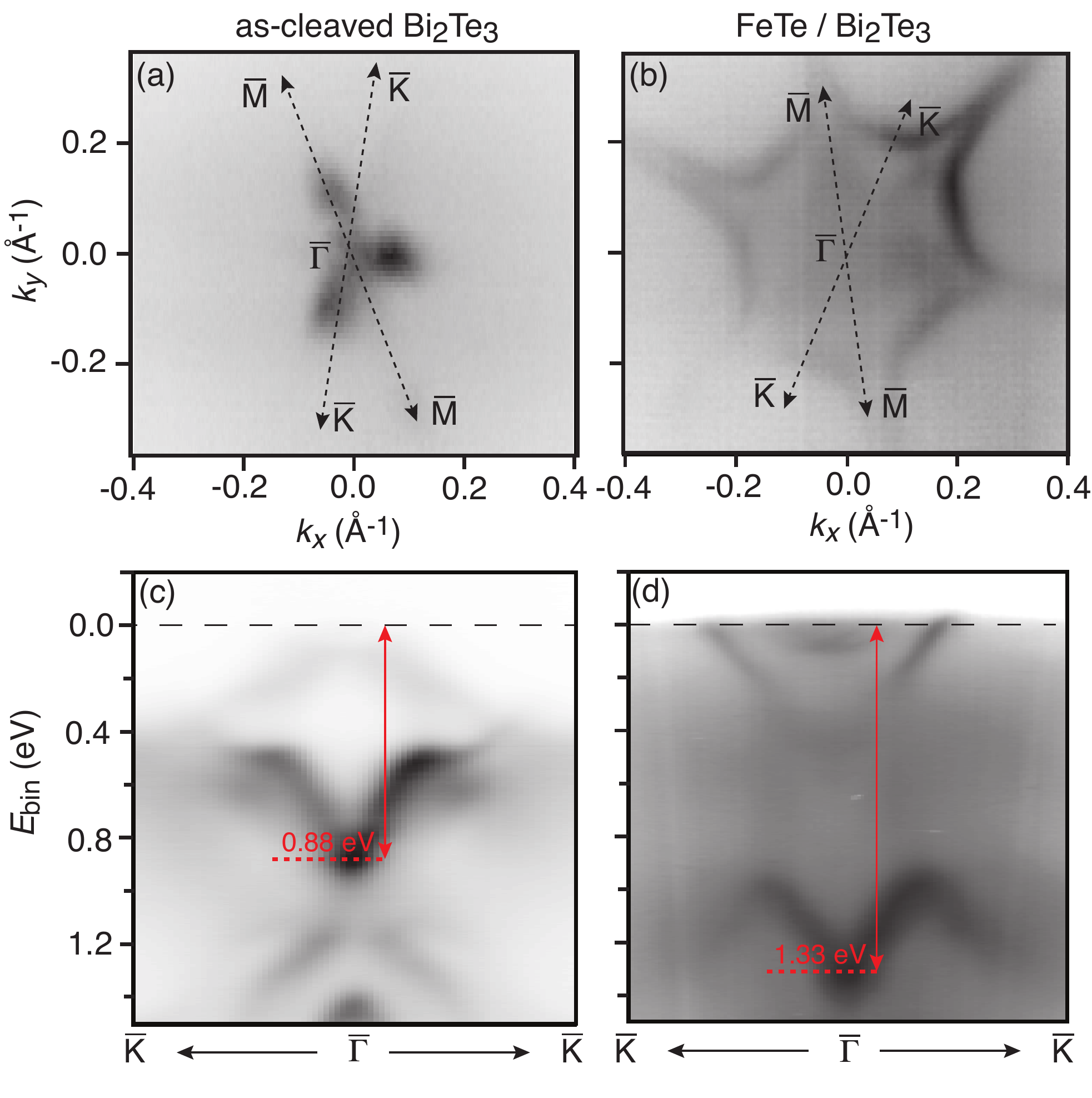}
		\caption{(Color online) (a)(b) Photoemission intensity in a 3~meV window centred on the Fermi energy and  (c)(d) dispersion along the $\bar{K}-\bar{\Gamma}-\bar{K}$ direction for as-cleaved Bi$_2$Te$_3$ and  thin film FeTe on Bi$_2$Te$_3$, respectively ($h{\nu} = 20$~eV). The red dashed line marks the bottom of the V-shaped surface state dispersion, obtained by a fit through an energy distribution curve in normal emission. The red arrows indicate binding energy change of this surface state. }
		
		\label{BiTe_FeTe_20eV}
	\end{center}
\end{figure}

An indirect way to address the doping of the FeTe film is via the change of the substrate's electronic structure \cite{Eich:2016}. While the states from the underlying bulk Bi$_2$Te$_3$ are not observable for ARPES data collected at a photon energy of 65~eV, they are dominating the spectra for $h\nu$=20~eV.  Fig. \ref{BiTe_FeTe_20eV} shows ARPES data taken at this photon energy for the as-cleaved surface of Bi$_2$Te$_3$ as well as for the thin FeTe film on Bi$_2$Te$_3$. Data from the clean surface in Fig. \ref{BiTe_FeTe_20eV}(a) and (c) reveals that the bulk samples have the valence band maximum (and the Dirac point of the topological surface state) very close to the Fermi level, and the surface state's dispersion thus in the unoccupied states. Covering the surface with FeTe induces a strong electron doping: Fig. \ref{BiTe_FeTe_20eV}(b) shows the strongly warped hexagonal Fermi contour of the topological surface state; the dispersion of the state is clearly visible in Fig. \ref{BiTe_FeTe_20eV}(d). Indeed, the doping of the surface is sufficiently strong to induce a  Rashba-split two-dimensional electron gas in the conduction band states \cite{Bianchi:2010ab,King:2011aa} and these give rise to the inner Fermi contour in Fig. \ref{BiTe_FeTe_20eV}(d). The features are not as well-defined as for alkali-doped pristine topological insulator surfaces \cite{Michiardi:2015aa}, presumably due to the surface being covered by the FeTe film. The pronounced substrate doping can be tracked by the intense V-like feature at higher binding energy which has been identified as a surface state of clean Bi$_2$Te$_3$ \cite{Michiardi:2014aa}. While the dispersion is slightly modified when the surface is covered by FeTe, it is still well-defined and shifted by $\approx$450~meV to higher binding energies. Overall, these observations support the notion of a charge transfer from the FeTe film to the substrate that leaves the film hole-doped. Note that the situation is similar to that reported for FeSe films on Bi$_2$Se$_3$ \cite{Eich:2016} where an almost equally large shift of the substrate bands and the topological state was found.

An important conclusion from the present section is that it is possible to interpret the photoemission data from FeTe  thin films, even in the presence of rotational domains. While not every point in the BZ becomes accessible, the hole pockets around $\bar{\Gamma}$ are clearly observed in higher BZs, where they are also well-separated for the different domains. Furthermore, a shift of the substrate's bands to higher binding energy is found, presumably causing a hole doping of the film.

\subsection{Doping of thin films}
\begin{figure}[b!]
	\begin{center}
		\includegraphics[width=1.0\linewidth]{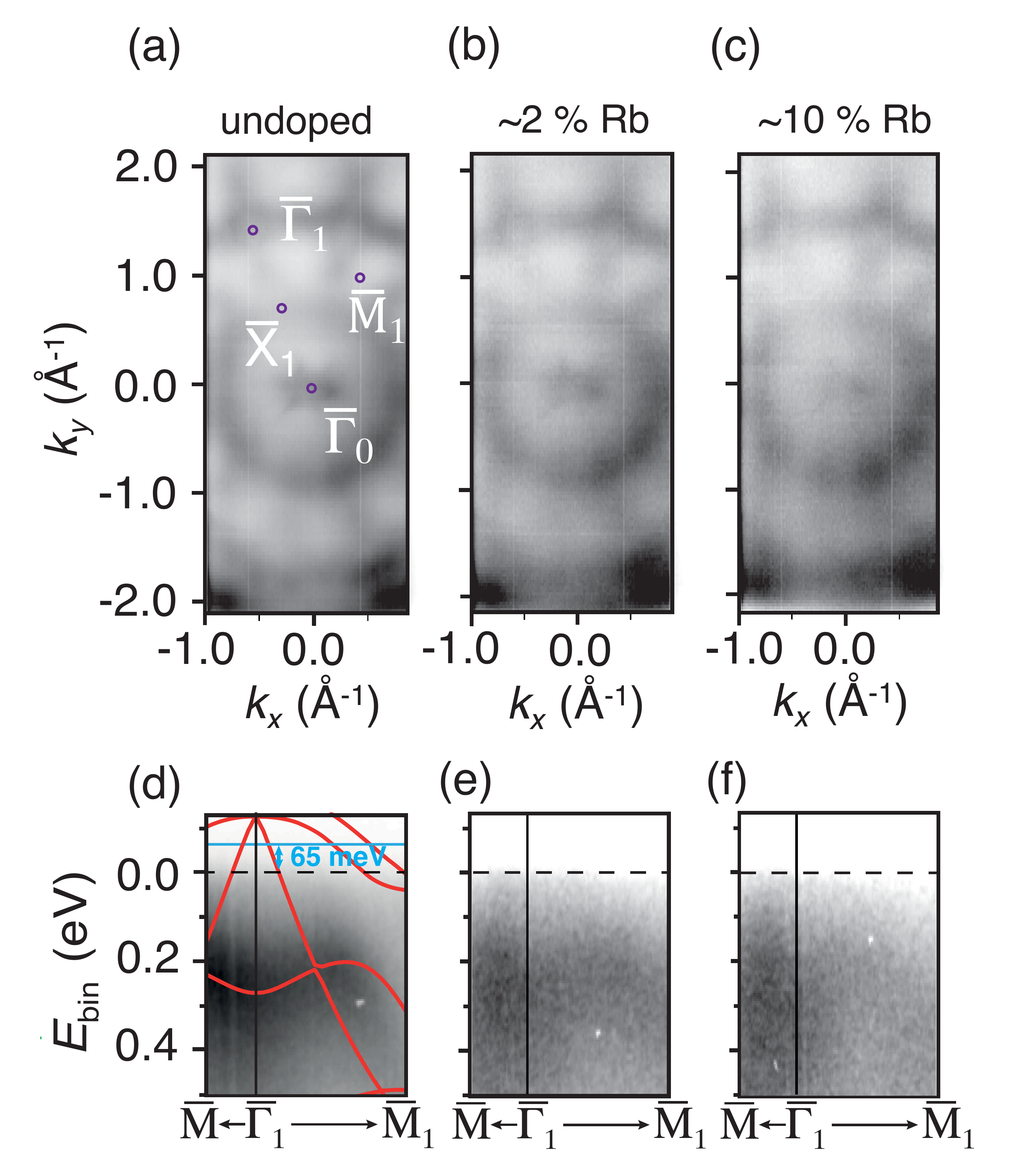}
		\caption{(Color online) Constant energy contour of thin film FeTe acquired at $h{\nu} = 65$~eV by integrating the spectral weight from a binding energy of +0.06 to -0.06~eV  of (a) the undoped thin film of FeTe on Bi$_2$Te$_3$, (b) $\sim$2~\% and (c) $\sim$10~\% Rb doped thin films.  High symmetry cuts for (d) the undoped (e) $\sim$2~\% and (f) $\sim$10~\% Rb doped thin films, respectively. The red solid lines are the first principles calculations in the  $\Gamma$-$M$-$X$-plane  and the light  blue line indicates the Fermi level of the first principle calculation.  } 
		
		\label{Thinfilm_doping}
	\end{center}
	
\end{figure}

Finally, we study the effect of an intentional local doping on the electronic structure of the FeTe films. Fig. \ref{Thinfilm_doping}(a)-(c) show the thin film's  photoemission intensity in a 120~meV window around the Fermi energy for different coverages of Rb atoms; Fig. \ref{Thinfilm_doping}(d)-(f) show the corresponding dispersion around $\bar{\Gamma}_1$. Surprisingly, even a high concentration of dopant atoms, with a coverage of $\sim$10~\%, does not lead to any noticeable changes of the Fermi contour. At the same time, all the features at higher binding energy become noticeably broader.

\begin{figure}[b!]
	\begin{center}
	\includegraphics[width=0.8\linewidth]{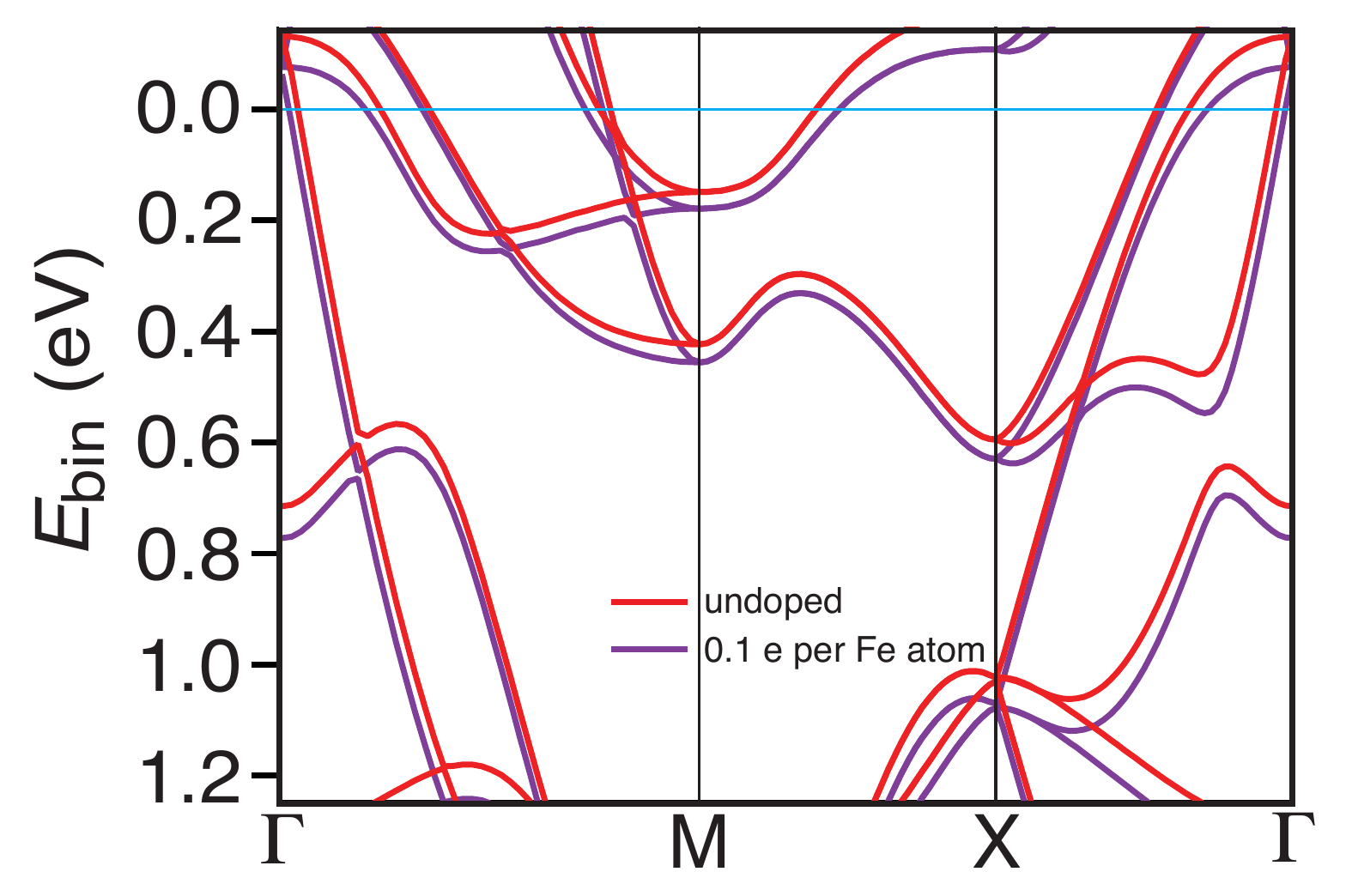}
		\caption{(Color online) Calculated band structure of an undoped and heavily doped  FeTe, showing only a minor doping-induced shift despite of the heavy doping.}. 
				\label{Thinfilm_doping_theory}
	\end{center}
	
\end{figure}

Alkali atoms are frequently used for surface doping and work function lowering due to the strongly ionic character of the bonding at low coverages (see e.g. Ref. \cite{Loptien2014}). It is therefore surprising that Rb adsorption does not lead to an observable doping here. In the simplest one-electron model, this behavior can be explained by the FeTe films' high density of states (DOS) at the Fermi level. Our calculations yield a DOS of approximately 5~eV$^{-1}$ per unit cell, i.e.
2.5~eV$^{-1}$ per Fe atom. In a rigid band model, even very strong electron doping on the order of 0.1~e per Fe atom merely results  in a doping-induced shift of the band structure on the order of 40~meV. This shift is obtained from DFT calculations considering an explicitly electron-doped cell with additional 0.1~e per Fe atom, as can be seen from Fig. \ref{Thinfilm_doping_theory}.  The figure also shows that the Fermi surface topology and band structure remains unchanged upon the amount of electron doping considered here. Clearly, electron correlation or lattice relaxation effects could alter this picture, particularly regarding details of the shape of Fermi surface. However, a high total density of states very similar to the DFT total DOS is also obtained in DMFT calculations \cite{Yin:2011}.

\section{Conclusions}

In conclusion, we have combined electronic structure investigations of cleaved bulk \FeTe crystals with predominantly single layer FeTe films grown on Bi$_2$Te$_3$. Using the results for the bulk crystals as a guide for the identification of electronic structure features in the thin film, the difficulty of the simultaneous presence of three rotational domains can be largely overcome. For this purpose, it turns out to be especially useful to use data taken around $\bar{\Gamma}_1$ in the second Brillouin zone. We find a close similarity of bulk and thin film electronic structure. Due to a strongly photon energy-dependent photoemission cross section for FeTe film and substrate states, we are able to observe the change in the substrate band structure upon growth of the FeTe film. We find an overall shift of the states to higher binding energy, suggesting a charge transfer from the film to the substrate.  Intentional doping of the FeTe films by Rb adsorption leads only to negligible shifts in the band structure, due to the films' high density of states at the Fermi level. Apart from yielding results for the thin film system, the approach outlined here should be applicable for the ARPES study of systems with multiple rotational domains.  

This work was supported by the Danish Council for Independent Research, Natural Sciences under the Sapere Aude program (Grant No. DFF-4002-00029), by VILLUM FONDEN via the Centre of Excellence for Dirac Materials (Grant No. 11744), the German Research Foundation via the DFG priority programme SPP1666 (grant no. WI 3097/2-2 and HO 5150/1-2) and by the ERC Advanced Grant ASTONISH (No. 338802). The work at Tulane  was supported by the US Department of Energy under Grant No. DE-SC0014208.

\end{document}